\documentclass[aps,prb,twocolumn,showpacs]{revtex4}

\usepackage{graphicx}
\usepackage{epsfig}
\usepackage{amsmath}
\usepackage{amssymb}
\usepackage{psfrag}
\usepackage{enumerate}

\newcommand{\bra}[1]{\langle #1 \vert}
\newcommand{\ket}[1]{\vert #1 \rangle}
\newcommand{\sub}[1]{_{\text{#1}}}
\newcommand{\super}[1]{^{\text{#1}}}
\newcommand{\str}{\operatorname{str}}
\newcommand{\opn}[1]{\operatorname{#1}}

\begin{document}
\title{Interface dependence of the Josephson-current fluctuations in short
mescoscopic superconductor/normal-conductor/superconductor junctions}
\author{T. Micklitz}
\affiliation{
Institut f\"ur Theoretische Physik, Universit\"at zu K\"oln,
Z\"ulpicher Strasse 77, 50937 K\"oln, Germany }
\date{\today}
\begin{abstract}
We discuss the dependence of the Josephson-current correlations in mesoscopic superconductor/normal-conductor/superconductor (SNS) devices on the transparency of the superconductor/normal-conductor interfaces. Focusing on short junctions we apply the supersymmetry method to construct an effective field theory for mesoscopic SNS devices which is evaluated in the limit of highly and weakly transparent interfaces. We show that the two-point Josephson-current correlator
differs by a universal factor of 2 in these two cases. 
\end{abstract}
\pacs{11}

\maketitle

\section{Introduction}
  One of the most promising directions in the field of 
  superconductivity in mesoscopic structures is provided by recent technological advances in
  hybrid superconductor/semiconductor technology. As an example we want to mention the very recently  
  observed quantization of the critical current in superconducting quantum point contacts \cite{quant}.
  An important aspect of this development is that for hybrid semiconductor
  structures both the system geometry and the chemical potential can
  be varied flexibly, which greatly facilitates the observation of
  {\it mesoscopic fluctuations}. In contrast, for ordinary
  metal heterostructures 
  only external magnetic fields effectively qualify as  averaging
  parameters, implying that the 
  dependence of fluctuation phenomena on the phase of the
  superconducting condensate is difficult to
  observe. Indeed, fabrication of such hybrid devices has advanced
  to the stage that fluctuation phenomena may be
  measured\cite{Takayanagi,Vegvar,Hartog,Kleinsasser,Morpurgo,Schapers} in a
  wide range of geometries and conditions.  
  
  The observable behavior of a two-dimensional electron gas (N), 
as realized in a semiconductor herterostructure, in contact with a superconductor (S)
is highly sensitive to the quality of the SN contact. In this work we discuss the SN interface dependence of the Josephson-current $J(\phi)$ through an electron
  gas, sandwiched between two superconductor terminals whose order
  parameters exhibit a phase difference $\phi$. A measurement of the
  fluctuation behavior ${\rm var\,}J(\phi)$ of this quantity has
  been attempted already by Takayanagi {\em et al}.\cite{Takayanagi}.
  As pointed out above, the use of a semiconductor heterostructure
  allows the use of an external gate voltage to tune fluctuations in
  the supercurrent, while keeping the phase difference {\em fixed}\cite{fn_otherexp}.
  For typical
  experiments\cite{Takayanagi,Kleinsasser} transmission tends to be
  weak due to the presence of the Schottky barrier. However, as noted
  in Ref.[\onlinecite{Takayanagi}], a theory of supercurrent fluctuations in
  the weak transmission regime has so far been missing. In this
  work, we fill this gap and compare the behavior of the
  supercurrent for weakly and highly transparent interfaces. (We remark that an 
  experimental realization of a
  crossover in the transition strength has already been attempted in
  Ref.[\onlinecite{Kleinsasser}].) 
  Specifically, we will compute the
  mesoscopic fluctuations of the current, ${\rm var}J(\phi)$ and, more
  generally, its {\it correlation profile}
\begin{eqnarray}
K(\phi_1,\phi_2) \equiv \langle J(\phi_1)J(\phi_2)\rangle
-\langle J(\phi_1)\rangle \langle J(\phi_2)\rangle.
\label{Kdef}
\end{eqnarray}
This latter function provides a wealth of information relating to the
design of the system, and it is clear that its experimental
determination -- along the lines of Ref.[\onlinecite{Takayanagi}] -- would
provide a sensitive test of the theory of mesoscopic fluctuations in
SN systems.  However, before proceeding, let us briefly put our
present analysis into the context of previous studies of the problem.

A pioneering work in the theory of supercurrent fluctuations is due to
Altshuler and Spivak\cite{A+Spivak}, who calculated the supercurrent
correlations by means of a diagrammatic perturbation theory.  Their
results assumed the limit of a long junction, so that $E_c \ll
\Delta$, where $E_c$ is the Thouless energy and $\Delta$ the
superconducting order parameter.  In this limit, the supercurrent
fluctuations are not universal: instead, the supercurrent variance is
proportional to $(e E_c/\hbar)^2$.

The applicability of the diagrammatic theory is limited, however, to
conditions under which the fluctuations of the  supercurrent exceed
its average. While such a situation may be reached by application
of a magnetic field, or in the presence of a glassy
structure in the N
region, it clearly is rather specialized.
Under more usual conditions, the essentially non-perturbative nature
of the proximity effect leads to a proliferation of diagrams whose
summation is practically impossible\cite{AST}.

A powerful alternative to the diagrammatic approach is provided by a
multiple scattering theory\cite{Been,Been2}. Here the property in
question is related to the scattering matrix of the N region, whose
statistical properties are known\cite{Been}.  The variance of the
supercurrent through the N region at fixed phase difference has been
calculated by Beenakker\cite{Been,Been2} using the scattering
approach. His results apply for a short junction $\Delta \ll
E_c$ with highly transparent SN interfaces.
In this case the fluctuations become universal: the supercurrent
variance is proportional to $(e\Delta/\hbar)^2$.

In this work, we apply the supersymmetry method\cite{Efetov} to construct an
alternative approach to the problem. 
A general feature of our formulation is its close alliance to the quasiclassical
approach\cite{Usadel,Eilenberger,VolkovRev,AST}, itself a traditional means of describing
the {\it mean} properties of SN systems. In particular, using the fact that the
stationary phase configurations of the field theory of
dirty\cite{fn_dirty} SN systems are determined by the quasiclassical
Usadel equations\cite{AST}, one can benefit from the huge
body of expertise on (ensemble averaged) Josephson-currents in
mesoscopic SN devices. The evaluation of
fluctuations around the Usadel mean field then leads to results for
the mesoscopic fluctuations of the current. 

To be specific, we apply our approach to the analysis of short
junctions with moderately weak coupling of the
superconductor condensate to the normal-metal ``quantum dot", which 
allows us to use random-matrix theory (RMT) to model the normal metal.
That is, we assume the hierarchy of energy scales $\Delta \ll E_g\ll E_c$, where $E_g\equiv \bar d g$ is
the inverse of the so-called dwell time\cite{Brouwer,Volkov}, 
$g\gg 1$ the normal-state conductance of the system, and $\bar d$ its mean level spacing.
(However, the application of the method to other regimes requires only technical rather than
conceptual modifications.)  Here onwards we put $\hbar =1$.

\section{Main result} 
Deferring an outline of the
technicalities of the analysis to the last part of this work, we here
merely anticipate that the correlator $K( \phi_1, \phi_2)$ is obtained by
a second-order perturbative expansion around the Usadel functional free
energy.  In this way we find, 

\begin{align}
\label{eq1}
  K_\Gamma(\phi_1, \phi_2)& = \frac{c_\Gamma e^2 \Delta^2}{\pi^2} \sin \phi_1 \sin
  \phi_2 \int_0^{\infty} dx_1 \int_0^{\infty} dx_2 \nonumber\\
  & \times \frac{\sqrt{x_1^2 +
      1} \sqrt{x_2^2 + 1}}{\sqrt{x_1^2 + \cos^2( \phi_1/2)}\sqrt{x_2^2
      + \cos^2( \phi_2/2)}} \nonumber 
      \\& \times \Big( \sqrt{x_1^2 + 1 } \sqrt{x_2^2 +
    \cos^2(\phi_2/2)} \nonumber \\
   & \quad  + \sqrt{x_2^2 +1 } \sqrt{x_1^2 +
    \cos^2(\phi_1/2)} \Big)^{-2}, 
    \end{align} where 
\begin{align}
\label{c}
c_\Gamma = \begin{cases} 1/2& \text{for highly transparent interfaces ($\Gamma=1$)}\\
1& \text{for weakly transparent interfaces ($\Gamma\ll1$)},
\end{cases}    
\end{align} and $\Gamma$ is the transmission coefficient characterizing the transparency of the SN interfaces; see below. While Eqs.~(\ref{eq1}) and (\ref{c}) are a new result for weakly transparent interfaces, 
current fluctuations of the moderately weakly coupled quantum dot with highly transparent interfaces have been the subject of an earlier scattering theory analysis in Ref.~\onlinecite{Been2}. Within this
approach both the diagonal contribution $\operatorname{var}J(\phi)=K(\phi,\phi)$\cite{Been2} and the straightforward
extension to the full correlation function $K( \phi_1,
\phi_2)$ are represented as a double integral over
eigenvalues of the transmission matrix, which despite its difference in form, is
numerically identical. The agreement with these earlier results provides a check on the consistency of the
field theory. Notice that the validity
of the result is limited to values of the phase outside the domain\cite{fn1} $|
\phi- \pi|<g^{-1}$, a fact that was pointed out earlier
in Ref.~[\onlinecite{Been2}]. The main result of this work is that the Josephson-current fluctuations for the case of weakly and highly transparent interfaces differ by a universal factor of 2. This factor 2, in fact, can be anticipated 
by the observation that for small phase differences $\phi$, one may establish
contact to the theory of universal conductance fluctuations (UCFs) through a disordered quantum dot--i.e., for $\phi \ll 1$
 \begin{align} 
  J(\phi) \approx  \frac{ge\Delta}{2}
  \phi, \end{align} where $g$ is the (dimensionless)
conductance through the disordered quantum dot. Using the fact that  for the
orthogonal ensemble\cite{12}

\begin{figure}[t] 
\centering \includegraphics[width=7.4cm]{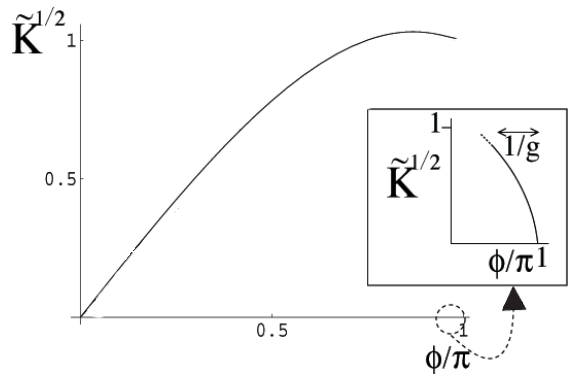}
\caption{\label{k1} Plot
of $\sqrt{\tilde{K}(\phi,\phi)}$ against phase, $\phi$, where
$\tilde{K}\equiv \pi^2/(c_\Gamma e^2\Delta^2) K$. The inset shows how
 $\sqrt{\tilde{K}(\phi,\phi)}$ falls to zero within a narrow window
around $\phi=\pi$.}
\end{figure}  

\begin{align}
 \operatorname{var}g  
&= \begin{cases}
  \frac{1}{4} & \text{ for high barriers } (\Gamma \ll 1) \\ 
  \frac{1}{8} & \text{ for transparent contacts } (\Gamma =1),
\end{cases} 
\end{align} one is led to the expectation ($\phi\ll 1$)

\begin{align} 
  \operatorname{var}J(\phi)= \begin{cases} \frac{e^2
      \Delta^2}{16} \phi^2 & \text{for high barriers } (\Gamma \ll 1) \\ 
    \frac{e^2 \Delta^2}{32} \phi^2 & \text{for transparent
      contacts } (\Gamma =1),\end{cases} \end{align} in agreement with Eqs.~(\ref{eq1}) and (\ref{c}). 
   
Figure~\ref{k1} shows the Josephson-current fluctuations $\sqrt{\operatorname{var}J(\phi)}=\sqrt{K(\phi,\phi)}$ as a function of $\phi$. We observe that the maximum falls to a fixed value $\phi\sim
2.7$. The strong increase of the fluctuations is a manifestation of the 
fact that $ \phi= \pi$ defines an instability of the Josephson action, in the vicinity of which minute
changes of the disorder/boundary geometry  trigger drastic
changes in the configuration-specific $J( \phi)$.

\section{Formalism} 
Having discussed our results for the 
SNS quantum dot, we next provide a brief outline  of the
formalism. 

Assuming $M$ channels coupling the disordered quantum dot to 
the superconductors (with each $M/2$ modes propagating to the left and to
the right, respectively\cite{fn2}) the Hamiltonian of the SNS junction is given by
\begin{align}  
   &H =  H\sub{d} + H^1_\Delta + H^2_\Delta + H^1\sub{c} +
   H^2\sub{c},
  \end{align} where ($i=1,2$)
\begin{align}   
&H\sub{d} = \sum_{\mu\nu} \ket{\psi_{\mu}}
   H\sub{dot}^{\mu\nu} \bra{\psi_{\nu}}, \\
 &H^i_\Delta = \sum_{a a'}
   \int dE \ket{\chi^i_a(E)} H\sub{BdG}^{i,a a'}
   \bra{\chi^i_{a'}(E)}, \\
 &H^i\sub{c} = \sum_{\mu a} \int dE \Big(
   \ket{\psi_{\mu}} W_i^{\mu a} \bra{\chi^i_a(E)} + \ket{\chi^i_a(E)}
   W_i^{\mu a} \bra{\psi_{\mu}} \Big). 
\end{align} $\{\ket{\psi_{\mu}}\}$ denotes a set of dot states (for further
  convenience we use the Nambu formalism; i.e.,~the
  states $\ket{\psi_{\mu}}$ comprise particle and hole degrees of
  freedom in a single object), and $\{ \ket{\chi^i_a(E)} \}$ a set of scattering states into superconductor $i$. The dot Hamiltonian is of the form $H\sub{dot} =
\sigma_3\super{ph} \otimes h$ where $h$ is drawn from the orthogonal Gaussian ensemble with the variance set by the spectrum width $\lambda$, $H\sub{BdG}^{i}$ is the Boguliubov-de Gennes Hamiltonian of superconductor $i$ with order parameters $\Delta_i=\Delta e^{i\phi_i}$, and the coupling matrices $W^i$ are characterized by the transmission coefficients\cite{fn3} $\Gamma$,
$W_i W_i^t = \frac{f(\Gamma)}{\pi} \lambda
\delta\super{op}_{\text{ch.},i}$, where $f(\Gamma) = \frac{2}{\Gamma} -
1 - \frac{2}{\Gamma} \sqrt{1 - \Gamma}$ and $\delta\super{op.}_{\text{ch.},i}$ describes a diagonal matrix with entries $1$ for the $M/2$ open channels connecting the dot to superconductor
$i$ and $0$ otherwise (We assume that the sets of dot states coupling to superconductors 
1 and 2 are disjoint.) The capacitance of the dot is assumed to be
sufficiently large that the Coulomb blockade can be ignored. 

To compute Josephson-current correlations we introduce a supersymmetric field integral representation of the free energy\cite{Efetov}:
\begin{align}
\label{k}
K(\phi_1,\phi_2)= \left(\frac{e}{\pi}\right)^2 \int d\omega_1\int d\omega_2 \frac{d^2  {\cal F}(\hat{\omega},\Phi)}{d\varphi_1d\varphi_2}|_{\varphi_1=\varphi_2=0},
\end{align} where 
\begin{align}
{\cal F}(\hat{\omega},\Phi)=\int {\cal D}Q e^{-S[Q]}
\end{align} with
\begin{align}
\label{fr2}
  S[Q]  = 
    \frac{M}{2}& \opn{Re} \str  \ln \Big( 1 \nonumber \\ 
   &  + \frac{
    f(\Gamma)}{\sqrt{\omega^2 + \Delta^2}} \Big[ \hat{\omega}
  \sigma_3\super{ph} + \Delta \sigma_2\super{ph} e^{ i \Phi
    \sigma_3\super{ph} \sigma_3\super{tr}}\Big]Q \Big).
\end{align} Here, $Q$ is a 16-dimensional matrix acting on the product of 
particle-hole (ph) space, a two component space (f) required to
distinguish between the two supercurrents, a boson-fermion (bf) space 
implementing the supersymmetric structure of the theory, and 
a time reversal (tr) space required to correctly describe the
behavior of the system under time reversal. Further, the symbol ``str"
denotes the supersymmetric generalization of the matrix trace, and  
$\hat\omega=\mbox{diag} (\omega_1,\omega_2)$ and $\Phi=\mbox{diag} (\phi_1+\varphi_1,\phi_2+\varphi_2)$  are diagonal matrices
in f-space where $\omega_{1,2}$ and $\phi_{1,2} \propto \openone\super{bf}$ are the two energy and phase
arguments entering the integral representation of the supercurrent  and the sources $\varphi_{1,2}=\opn{diag}(\varphi_{1,2}, 0 )\sub{bf}$ break supersymmetry.

Referring for a detailed discussion of the derivation of Eqs.
(\ref{k})-(\ref{fr2}) and of the internal structure of the matrix $Q$ to
Ref.[\onlinecite{AST}], we here merely recapitulate, that it originates from the coherent state representation of the free energy along the standard procedure, including
the generalization of the partition function to a supersymmetry formulation, RMT ensemble average, Hubbard-Stratonovich transformation and expansion in Goldstone modes around the metallic saddle point $\sigma_3\super{ph}$.

Notably,
the nonlinear constraint $Q^2=\openone$ indicates that the matrix $Q$
represents a generalization of the Green function $g$ central to the
quasiclassical approach. Indeed, it is straightforward to see that a
variation of Eq.  (\ref{fr2}) (under the restriction $Q^2=\openone$)
leads to
\begin{align}
\label{Usq2}
 \Big[ \bar{Q}, \omega \sigma_3\super{ph} + \Delta \cos \Phi
\sigma_2\super{ph} \Big] = 0,
\end{align} i.e., an equation which, upon identification of the matrix $Q$
with the quasiclassical Green functions, immediately is recognized as the
zero-dimensional limit of a Usadel equation (extended, however, to a
larger structure accommodating more than one impurity averaged
observable.) Void of elements coupling between the two observables
(f-space), Eq. (\ref{Usq2}) admits a block-diagonal solution $\bar{Q}= \mbox{diag}(g_1,g_2) \otimes \openone^{\rm bf} \equiv R \sigma_3^{\rm ph}
R^{-1}$, where 
\begin{align}
g_i=\frac{1}{\sqrt{\omega_i^2+\Delta^2}}\left( \omega_i \sigma_3\super{ph} + \Delta \cos \Phi_i \sigma_2\super{ph}\right)
\end{align} are the quasiclassical Green functions
computed for phase difference $\Phi_{1,2}$ and the second
representation expresses the solution as a rotation away from the
metallic reference point, $\sigma_3^{\rm ph}$. The ensemble-averaged
Josephson-current may now be calculated from the saddle-point action

\begin{align}
\label{mf}
S[\bar{Q}] = \frac{M}{4} \str \ln \Big( 1 + f^2(\Gamma) +
2 f(\Gamma) \frac{\sqrt{ \omega^2 + \Delta^2 \cos^2
    \Phi}}{\sqrt{\omega^2 + \Delta^2}} \Big), 
    \end{align} and one obtains for highly transparent interfaces 
\begin{align}
\label{mfc1}
  J(\phi) =& \frac{e M}{2\pi} \Delta \sin \phi
  \int_0^{\infty}dx \nonumber\\
  & \frac{1}{\sqrt{x^2 + 1} +\sqrt{x^2 + \cos^2
      \phi/2}} \frac{1}{\sqrt{x^2 + \cos^2 \phi/2}},
\end{align} while for weakly transparent interfaces

\begin{align}
\label{mfc2}
 J(\phi) =& \frac{e M \Gamma}{4 \pi} \Delta \sin\phi
  \int_0^{\infty}dx \frac{1}{\sqrt{x^2 + \cos^2 \phi/2}}
  \frac{1}{\sqrt{x^2 + 1}}.
  \end{align} Equations÷(\ref{mfc1}) and (\ref{mfc2}) coincide with quasiclassical results\cite{Volkov,K+Luk,Brouwer,Been}.

{\it Josephson-current correlations} can now be explored by
defining\cite{AST} 
\begin{eqnarray}
Q= R T \sigma_3^{\rm ph}  T^{-1} R^{-1},
\label{Qparam}
\end{eqnarray}
where the generalized rotation matrix $T$ describes
fluctuations around the Usadel mean field and the
free energy $ {\cal F}(\hat{\omega},\Phi)= \int {\cal D}T \exp(-S[Q])$. 

Going beyond the mean-field level by substituting the generalized
parametrization, Eq.~(\ref{Qparam}), into the field integral, one finds
that fluctuations around $T=\openone$ are penalized by a parameter $
\cos(\phi_i/2) g$ which is much larger than unity, {\it unless}
$|\phi-\pi|<g^{-1}$ lies in the anomalous window discussed
above. While in the latter case the correlator, Eq.~(\ref{k}), has to
be evaluated by full integration over the manifold of $T$'s (cf. the
analogous situation with computing the fine structure of spectral
correlation functions on energy scales smaller than the single
particle level spacing\cite{Efetov}), in general it is sufficient to
perturbatively expand the matrix $T=\openone +  W + {1\over 2} W^2 +
\dots$ to second order around unity.  Expansion of the logarithm then leads to the 
 Gaussian actions for highly and weakly transparent interfaces,

\begin{align} 
  S_{\Gamma\ll1}^{(2)}[W] &=  \frac{\Gamma M}{4} \str \left[\Pi_{\hat{\omega}}(\Phi) W^2\right], \\
%  \end{align} 
%\begin{align}
  S_{\Gamma=1}^{(2)}[W] &= \frac{M}{4} \str \Big[ W^2 \nonumber\\
  &+ \left( \frac{1}{1+\Pi_{\hat{\omega}}(\Phi)} \frac{\Delta}{\sqrt{ \hat{\omega}^2+\Delta^2}} 
  \sin\Phi\sigma_2\super{ph}\sigma_3\super{tr}W \right)^2\Big], 
      \end{align} where we introduced $\Pi_{\hat{\omega}}(\Phi) := \frac{\sqrt{ \hat{\omega}^2 + \Delta^2 \cos^2
      \Phi}}{\sqrt{ \hat{\omega}^2 + \Delta^2}}$. Performing the twofold derivative with respect to the sources followed by Gaussian integrals over the $W$'s one obtains the final result, Eqs.~(\ref{eq1}) and (\ref{c}). 

\section{Generalization} 
While the outline above was specific to the case
of a spatially structureless quantum dot, the generalization to more
complex geometries is straightforward. Under conditions where the
spatial variation of the quasiclassical Green function is no longer
negligible, Eq. (\ref{fr2}) generalizes to the action of a matrix {\it
  field} $Q$. For example, for an extended diffusive N region (i.e., for $E_c
< \Delta,E_g$) and weakly transparent interfaces, $S[Q] = -(\pi D \nu/ 8) \int d^dr{\,\rm
  str}(\partial Q\partial Q) + S_\omega[Q] + S_c[Q]$, assumes the form
of a generalized diffusive $\sigma$ model, \cite{AST} where
$S_\omega[Q] = -(\pi \nu/2)\int d^dr {\,\rm str}(Q\hat \omega
\sigma_3^{\rm ph})$
and $S_c[Q]$ describes the coupling to the superconductor. Similarly,
for a (nearly) clean system, $S[Q] = -v_F \int d^d r dn {\,\rm
  str}(T\sigma_3^{\rm ph} {\bf n}\cdot \partial T^{-1})
+S_\omega[Q]+S_c[Q]$, becomes the free energy of the ballistic
$\sigma$-model\cite{Muz}. In complete analogy to our discussion above, the
variation of these actions leads to the general Usadel\cite{Usadel} and
Eilenberger\cite{Eilenberger} 
equations, respectively. To compute Josephson-current
correlations, one again employs the parametrization, Eq.~(\ref{Qparam}),
where, however, both $R$ and $T$ are field configurations with
nonvanishing spatial variation. By evaluating the action of a
generalized quadratic variation $T=\openone+W+{1\over 2}W^2$ and
subsequent computation of the correlator, Eq.~(\ref{k}), one can then,
in principle, obtain the Josephson-current correlations of any
SN system category (if amenable to the approximation schemes of
quasiclassics.) However, in cases where the solution of the
Usadel and Eilenberger
equations displays a complex spatial profile, the concrete computation
of the generalized Gaussian integrals over $W$ can be
cumbersome. Under these conditions one expects the result for the
current correlations to be less universal (e.g., dependent on the
system geometry, disorder concentration, etc.) than in the quantum dot
case discussed above. 

\section{Summary}
In summary, we showed that the mesoscopic Josephson-current fluctuations
through a weakly coupled disordered quantum dot differ by an universal factor of 2 for the cases of weakly and highly transparent interfaces. In our calculations we applied a supersymmetric field-theoretical approach. Although we concentrate on the case of short junctions (and use RMT to model the normal metal), this method allows for a microscopic derivation capable to cover a wide range of parameters characterizing the SNS junction. The observation of the Josephson-current correlations should be feasible using semiconductor technology. 
 
 I am grateful to A.~Altland for driving my attention to this subject and for many helpful discussions.
 Furthermore, I acknowledge financial support of the SFB/TR~12 of the Deutsche Forschungsgemeinschaft.


\begin{references}
\bibitem{fn_dirty} An SN system is  ``dirty" if the scattering mean free
  path is smaller than both, the size of the normal region
  and the coherence length $ \Delta/v_F$ ($v_F$: Fermi velocity). The
  complementary case of quasi-clean SN systems is described by a
  ballistic variant of the field theory whose equations of motion are
  the Eilenberger equations~\protect\cite{Muz,Eilenberger}.
\bibitem{fn2}
For simplicity we assume that the number of modes coupling to superconductor S$_1$ and S$_2$ is the same. This simplification has also been made in Ref.[\onlinecite{Been2}].
\bibitem{fn3}
$\Gamma$ are the transmission eigenvalues of the S-matrix.
\end{references}
\end{document}